\documentclass[journal=jacsat,manuscript=article]{achemso}

\usepackage{chemformula} 
\usepackage[T1]{fontenc} 
\usepackage{subcaption}

\author{Ewout van der Veer}
\email{ewout.van.der.veer@rug.nl}
\affiliation[University of Groningen]
{Zernike Institute for Advanced Materials, Nijenborgh 4, 9747AG Groningen, The  Netherlands}
\author{M\'onica Acuautla}
\email{m.i.acuautla.meneses@rug.nl}
\affiliation[University of Groningen]
{Engineering and Technology Institute Groningen (ENTEG), Nijenborgh 4, 9747AG Groningen, The  Netherlands}
\author{Beatriz Noheda}
\email{b.noheda@rug.nl}
\affiliation[University of Groningen]
{Zernike Institute for Advanced Materials, Nijenborgh 4, 9747AG Groningen, The  Netherlands}

\title[An \textsf{achemso} demo]
  {Ferroelectric  \ch{PbZr_{1-x}Ti_xO_3} by ethylene glycol-based chemical solution synthesis}

\abbreviations{IR,NMR,UV}

\begin{document}

\begin{abstract}
We have investigated a water-stable sol-gel method based on ethylene glycol as a solvent and bridging ligand for the synthesis of ferroelectric lead zirconate titanate in bulk and thin film forms. This method offers lower toxicity of the solvent, higher stability towards atmospheric moisture and a simplified synthetic procedure compared to traditional sol-gel methods. Ceramic pellets of Nb-doped lead zirconate titanate (PNZT) in the rhombohedral phase were produced with high density and good piezoelectric properties, comparable to those reported in the literature and those found in commercial piezoelectric elements. In addition, a nine-layer thin film stack was fabricated from the same sol by spin coating onto platinized silicon substrates. The films were crack-free and showed a dense perovskite grain structure with a weak (111) orientation. Piezoelectric measurements of the film showed a piezoelectric coefficient comparable to literature values and good stability towards fatigue. 
\end{abstract}
\section{Introduction}
Sol-gel methods are commonly used for the fabrication of oxide materials with a wide range of functionalities. These methods involve the synthesis of a precursor solution (known as `sol') containing oligomeric chains of metal ions and oxygen atoms. Treatment of the sol, for example by the addition of water or by heating, causes the formation of a continuous metal-oxygen network, leading to gelation of the sol. The sol may be processed into a variety of products, such as bulk powders (by simply heating the gel), thin films by deposition of the sol onto a substrate (\textit{e.g.} by spin coating) or a multitude of other forms.\cite{Danks2016, Bassiri-Gharb2014}  Employing a sol-gel-type synthesis for the production of oxide materials facilitates the control of composition and doping, making high homogeneity and short fabrication cycles possible.\cite{Danks2016} Furthermore, when used to fabricate thin films, it can give rise to smooth films covering a large surface area with a wide range of film thicknesses up to several micrometers.
 
One material which may be produced using a sol-gel method is the well-known lead zirconate titanate solid solution (\ch{PbZr_{1-x}Ti_xO_3}, also known as PZT), which is ferroelectric and, thus, piezoelectric, allowing for its use as sensors and actuators. The PZT composition with x=0.48 lies at a phase boundary between two different crystal structures with tetragonal (for Ti-rich compositions) and rhombohedral (for Zr-rich compositions) symmetries, where monoclinic structures have been observed\cite{Noheda1999}. At this boundary, known as as the morphotropic phase boundary (MPB), the piezoelectric coefficients are maximized.\cite{Jaffe1971} The piezoelectric parameters of PZT can be further improved through chemical doping with elements such as niobium (\ch{PbNb_y(Ti_xZr_{1-x})_{1-y}O_3}  or PNZT).\cite{Damjanovic1999} 

Traditional sol-gel methods used for the production of thin films of PZT, first reported by Budd, Dey and Payne\cite{Budd1985}, make use of the highly toxic 2-methoxyethanol as a solvent and alkoxides and acetates as the precursors for lead, zirconium and titanium. These methods rely on hydrolysis and condensation reactions of the alkoxide precursors to form a polymeric network of metal-oxygen-metal bonds. These methods use water for the initiation of the hydrolysis reaction. Hence, sols produced using such an approach tend to be sensitive to the presence of water.\cite{Danks2016} As a result, these sols require storage and processing in an oxygen-free and water-free environment, such as a glovebox. 

More recently, there has been an interest in the development of chemical solution deposition (CSD) methods which are not based on hydrolysis-condensation reactions, instead relying on different types of reactions.\cite{Danks2016, Niederberger2007, Vioux1997, Debecker2012} One example of such a non-aqueous CSD method is based on ethylene glycol as bridging ligand and common alkoxides and acetates as reagents. This method was reported to be nontoxic, more stable to atmospheric moisture and have a more straightforward synthesis procedure.\cite{De-Qing2007} However, an investigation of the ferroelectric and piezoelectric properties of materials derived from this CSD method has, to our knowledge, not been reported.

We have studied the properties of both bulk and thin film products fabricated using the ethylene glycol-based CSD method\cite{De-Qing2007} . We show that multilayer stacks of thin films can be produced without cracks, voids or parasitic phases by carefully designing the deposition and heat treatment procedures, despite the presence of a large amount of organic material in the as-deposited film, which is commonly known to reduce film quality.\cite{Damjanovic1997} The piezoelectric behavior of the bulk and thin film products are comparable to reported values for films of similar characteristics. Finally, we have investigated the sensitivity of the solution to moisture. 

\section{Results and discussion}

\label{sec:res}
Properties of the sol as well as structural and ferroelectric properties of the PNZT films and bulk ceramic pellets have been investigated. 

\begin{figure}[ht]
\centering
\includegraphics[width=0.7\columnwidth]{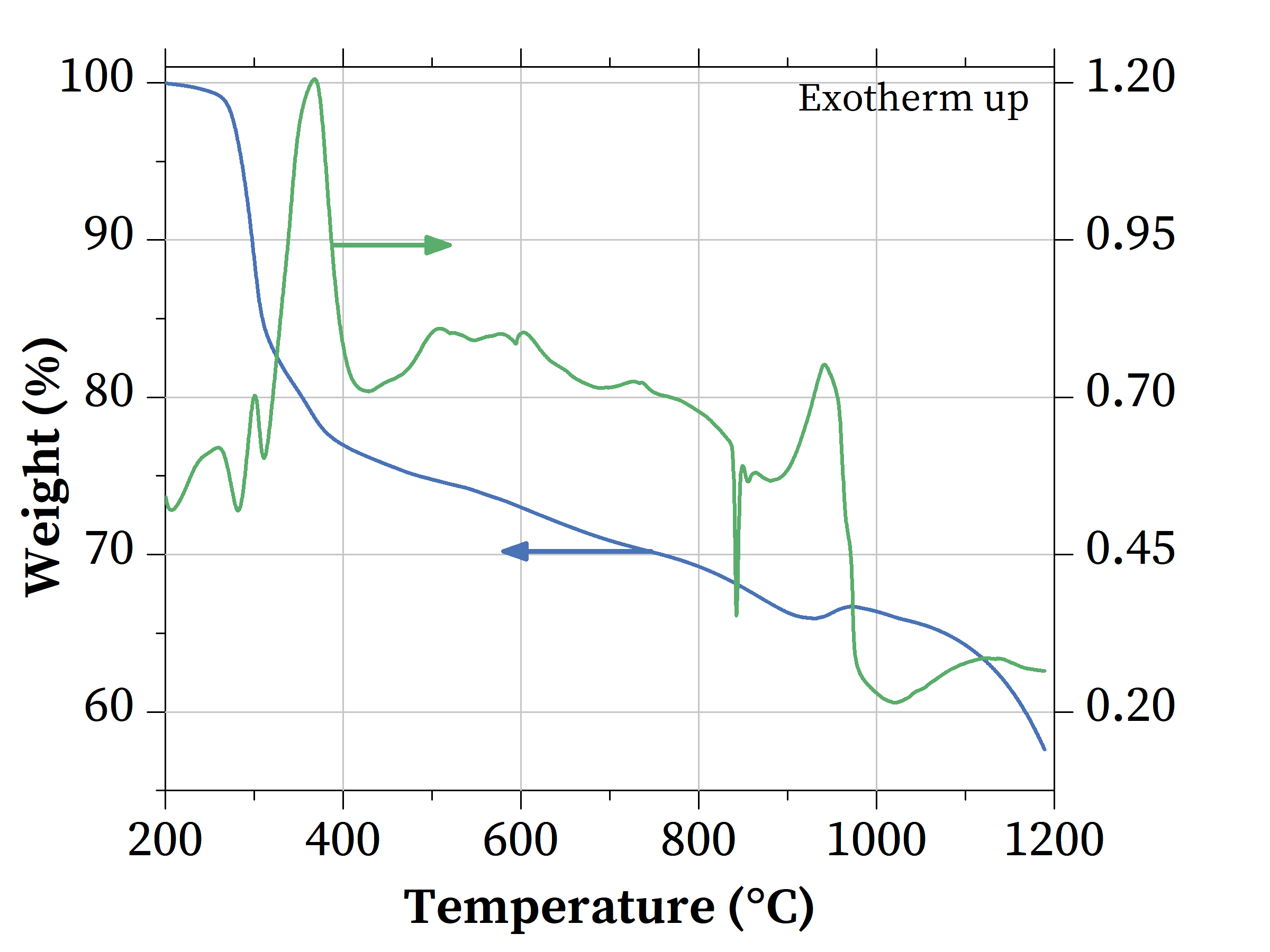}
\caption{DTA and TGA traces of the sol dried at 230\textcelsius.}
\label{fig:dsc}
\end{figure}

Figure \ref{fig:dsc} contains plots of the differential thermal analysis (DTA) and thermo-gravimetric analysis (TGA) data collected from the sol dried at 230\textcelsius\ on a hotplate. Initial weight loss occurs around 300\textcelsius\ and is associated with a peak in the DTA trace. This peak corresponds to loss of ethylene glycol groups. Further weight loss occurs between 320\textcelsius\ and 400\textcelsius, corresponding to a large exothermic peak in the DTA trace. This peak is presumably the result of the removal of remaining organic material and the onset of crystallization of PNZT.\cite{Livage1994, Tu1995a} A total weight loss of approximately 23\% was observed up to 400\textcelsius. The final peak present at 842\textcelsius\ is possibly due to melting of lead oxide in the sample. These results are in rough correspondence with those reported in the literature using a similar ethylene glycol-based solution deposition method\cite{Livage1994}. 

The sensitivity of the sol to the presence of water (for example, from the atmosphere), was determined by directly adding various concentrations of water to 1 mL of the sol. The sol was left at room temperature in a dark location for over a month, yet no gelation occurred even in sols to which 10 vol.\% water was added. After a month, some gelation was observed, but there was no correlation between the gelation time and the concentration of water that was added to the sol. Sols with up to 5 vol.\% water were used to produce pellets as described in the `Materials and Methods' section below. These pellets were analyzed by x-ray diffraction and scanning electron microscopy to assess the influence of the addition of water to the sol on the structural and microstructural properties of the product. No trends could be discerned in either the structural or the microstructural properties as the concentration of water was increased. These results show that the ethylene glycol-based sol is highly stable towards moisture.

\subsection{Bulk}
PNZT pellets were produced from the PNZT sol with a 20\% excess of lead precursor. The sol was dried at 230\textcelsius\ and pyrolyzed at 420\textcelsius. Pellets with a nominal diameter of 10 mm were pressed from this powder at 6.4 ton/cm$^2$. 
One pellet was sintered at 800\textcelsius\ for 2 hours. X-ray diffraction analysis of this pellet suggested that the pellet was in the perovskite phase. Nevertheless, the peak splittings expected for either the rhombohedral or tetragonal phases of PNZT were not present and no ferroelectric behavior was measured in this pellet. A second pellet was sintered at 1200\textcelsius\ for 2 hours. After sintering, the pellet had a diameter of 8.16 mm, a thickness of 1.39 mm and a density of 6936 $\mathrm{kg/m^3}$, that is 86.9\% of the theoretically predicted density\cite{Jaffe1954}. 

\begin{figure}[htb]
\centering
\includegraphics[width=0.7\columnwidth]{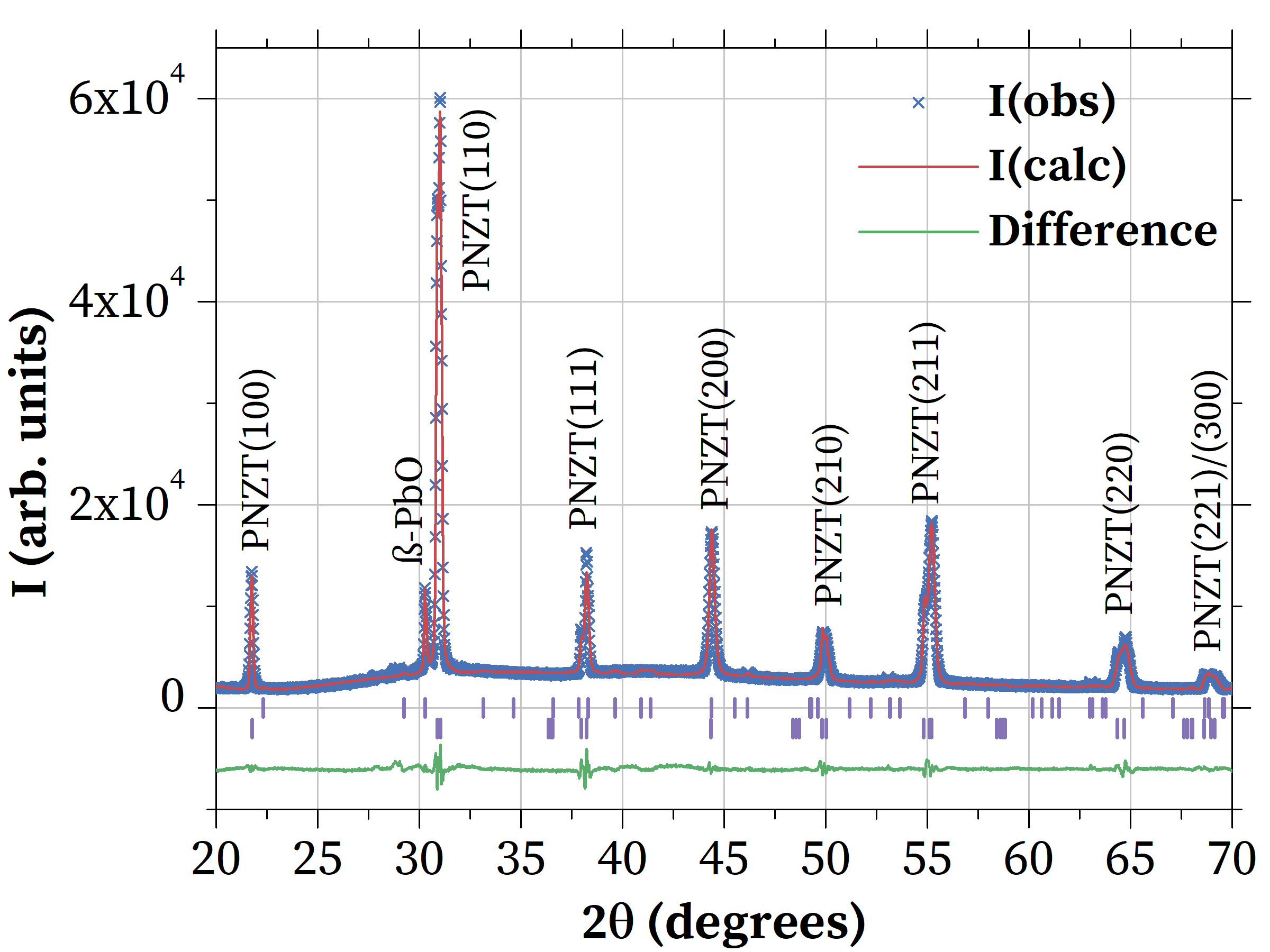}
\caption{X-ray diffraction pattern of the pellet and a fit of the profile using rhombohedral PNZT and $\mathrm{\beta}$-PbO phases.}
\label{fig:xrdbulk}
\end{figure}

\begin{figure}[htb]
\centering
\includegraphics[width=0.7\columnwidth]{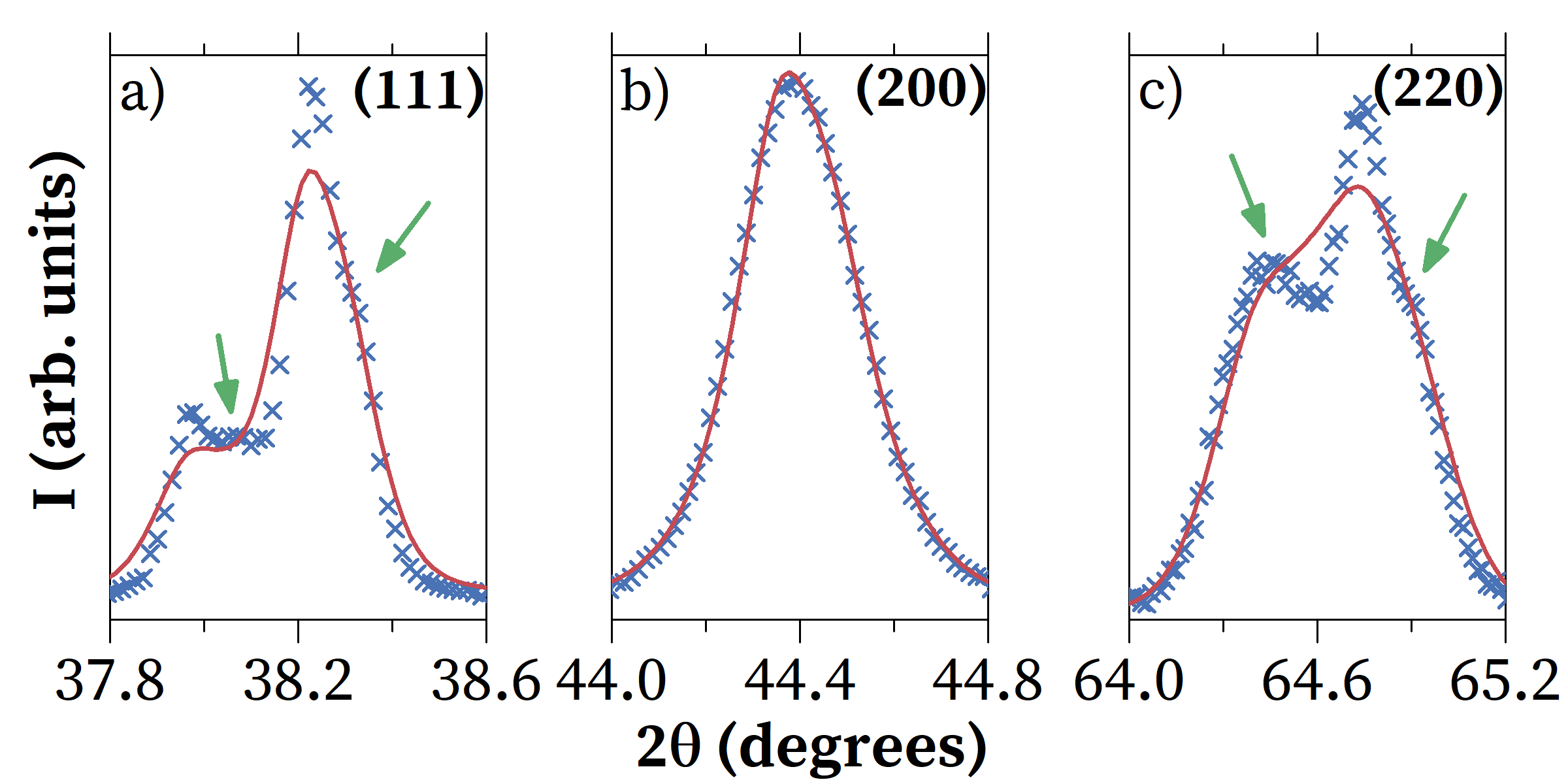}
\caption{Blow-ups of the (a) (111), (b) (200) and (c) (220) peaks of the pattern in figure \ref{fig:xrdbulk}. Peaks originating from a tetragonal or monoclinic phase are indicated using green arrows.}
\label{fig:xrdblowup}
\end{figure}

Figure \ref{fig:xrdbulk} shows an x-ray diffraction pattern of the pellet after sintering at 1200\textcelsius. A good fit of this pattern was obtained using a combination of a rhombohedral PNZT phase and a $\mathrm{\beta}$-PbO phase. This indicates that the excess of lead precursor in the sol is too high, leading to the formation of an impurity phase. Nevertheless, good quality PNZT pellets could not be obtained using a lower lead excess. Blown-up versions of the (111), (200) and (220) peaks of the pattern are shown in figure \ref{fig:xrdblowup}. The splitting of the peaks indicates that the material is in a mostly rhombohedral phase, with a small admixute of a tetragonal or possibly a monoclinic phase\cite{Noheda1999, Noheda2000, Noheda2000b}. Hence, the material is approaching the morphotropic phase boundary between the rhombohedral and tetragonal phases known to present the best piezoelectric response.\cite{Jaffe1971} The slight deviation from the exact composition at the morphotropic phase boundary may result from the loss of titanium precursor during synthesis due to its high reactivity with atmospheric moisture or impurity of the precursor itself.

\begin{figure}[htb]
\centering
\includegraphics[width=0.7\columnwidth]{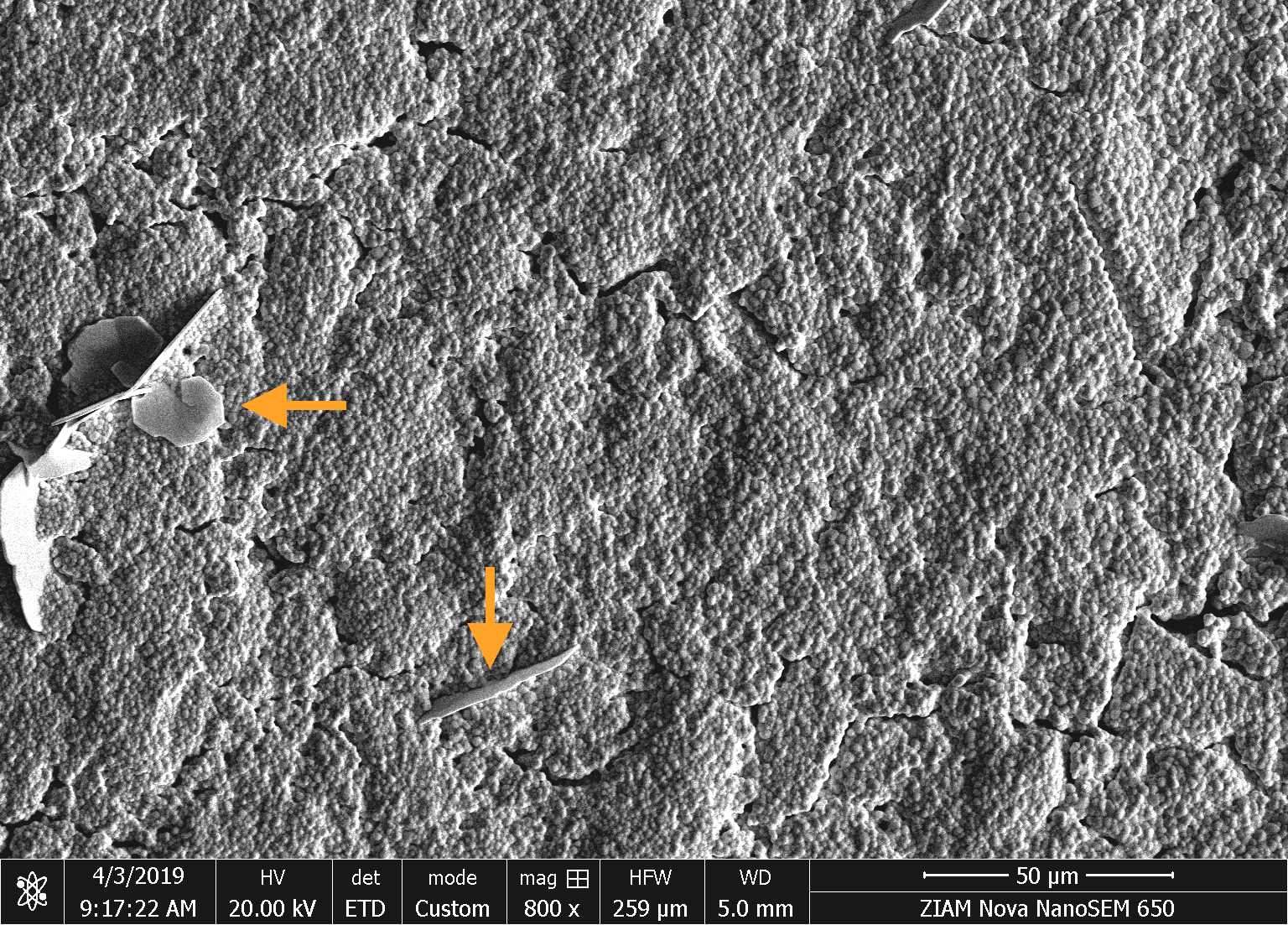}
\caption{Scanning electron microscopy image of a pellet sintered at 1200\textcelsius\ showing PNZT grains and larger lead oxide crystals (orange arrows).}
\label{fig:pelletsem}
\end{figure}

Scanning electron microscopy (SEM) of the pellet (figure \ref{fig:pelletsem}) shows a dense grain structure with PNZT grains of 500-1000 nm. Additionally, large, plate-like crystals are present in the PNZT matrix. These crystals were determined to be lead oxide by energy dispersive spectroscopy (EDS), confirming the presence of a lead oxide phase in the pellet. 

\begin{figure}[ht]
\centering
\includegraphics[width=0.7\columnwidth]{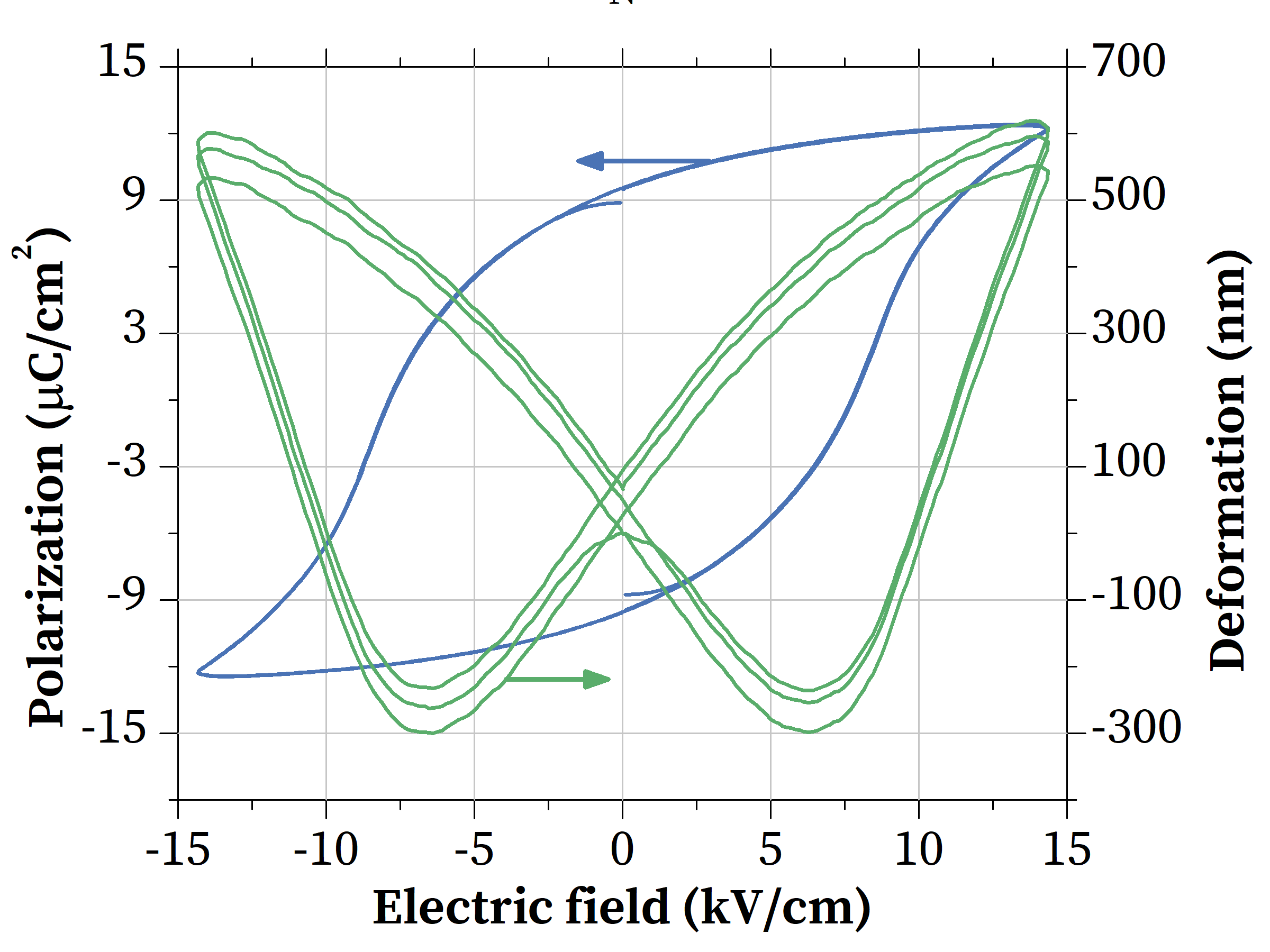}
\caption{Polarization and strain loops of sol-gel derived PNZT pellet sintered at 1200\textcelsius.}
\label{fig:piezopellet}
\end{figure}

The pellet was poled in a silicone oil bath at 100\textcelsius\ with an electric field of 29 kV/cm for 30 minutes to align the dipoles in the material. Ferroelectric property measurements of the bulk ceramic pellet were performed, yielding the polarization-electric field hysteresis loops and strain-electric field ("butterfly") loops expected for a ferroelectric, as displayed in figure \ref{fig:piezopellet}. The remnant polarization measured for this pellet is P$_r$= 9.5 $\mathrm{\mu C/cm^2}$, the coercive field E$_c$= 7.78 kV/cm and the longitudinal piezoelectric coefficient d$_{33}$= 441 pm/V. The d$_{33}$ coefficient obtained here is compared to literature values in table \ref{tab:piezocomp}. Our PNZT pellet has piezoelectric properties in line with those found in literature, even competing with commercially available piezoelectric elements. We expect that the ferroelectric and piezoelectric parameters can be further increased by bringing the composition closer to the morphotropic phase boundary and by improved densification of the pellet by, for example, hot pressing. This work shows that the ethylene glycol CSD method is capable of producing a high-quality material despite the simplicity of the method. 

\begin{table}[ht]
    \centering
    \begin{tabular}{l l l l}
        $d_{33}$ \textit{(pm/V)} & \textit{Doping element} & \textit{Production method} & \textit{Reference} \\
        \hline
        441 & Nb & CSD & This work\\
        \hline
        500 & Commercial & n.a. & \citet{Hinterstein2011}\\
        475 & Commercial & n.a. & \citet{Xu2013}\\
        420 & Undoped & Sol-gel & \citet{Sharma2001}\\
        155 & Undoped & Wet chemical & \citet{Choy1997}\\
        300 & Undoped & Wet chemical & \citet{Guiffard1998}\\
        569 & La & Sol-gel & \citet{Shannigrahi2004}\\
        269 & Nd & Sol-gel & \citet{Shannigrahi2004}\\
        325 & La & Wet chemical & \citet{Sahoo2013}\\
        236 & \ch{BiFeO3}/\ch{BaCu_{0.5}W_{0.5}O3}/\ch{CuO} & Solid-state & \citet{Dong1993}\\
        338 & La/Nb & Solid-state & \citet{Singh2006}\\
        520 & Sr/Nb & Solid-state & \citet{Zheng2001}\\
        255 & Nb & Solid-state & \citet{Garcia2007}\\
        \hline
    \end{tabular}
    \caption{Comparison of the longitudinal piezoelectric coefficient of the PNZT pellet fabricated using the ethylene glycol CSD method with literature values. }
    \label{tab:piezocomp}
\end{table}

\subsection{Thin films}
A nine-layer PNZT film was produced from the 1.5 M PNZT sol by spinning at 5000 rpm followed by drying on the hotplate, with pyrolysis and annealing steps performed after every third layer. During heat treatment of these films, lead can be lost through evaporation at the film surface and through diffusion into the silicon substrate. This leads to the formation of a layer of lead-deficient pyrochlore phase at the film surface or at the film-electrode interface. An excess of lead precursor can be added to the PNZT sol to compensate for this loss. However, too large an excess can cause the formation of voids in the film due to evaporation of the excess lead species. Therefore, careful control of the excess is required. 

\begin{figure}[ht]
\centering
\begin{subfigure}[h]{\columnwidth}
\centering
	\includegraphics[width=0.7\columnwidth]{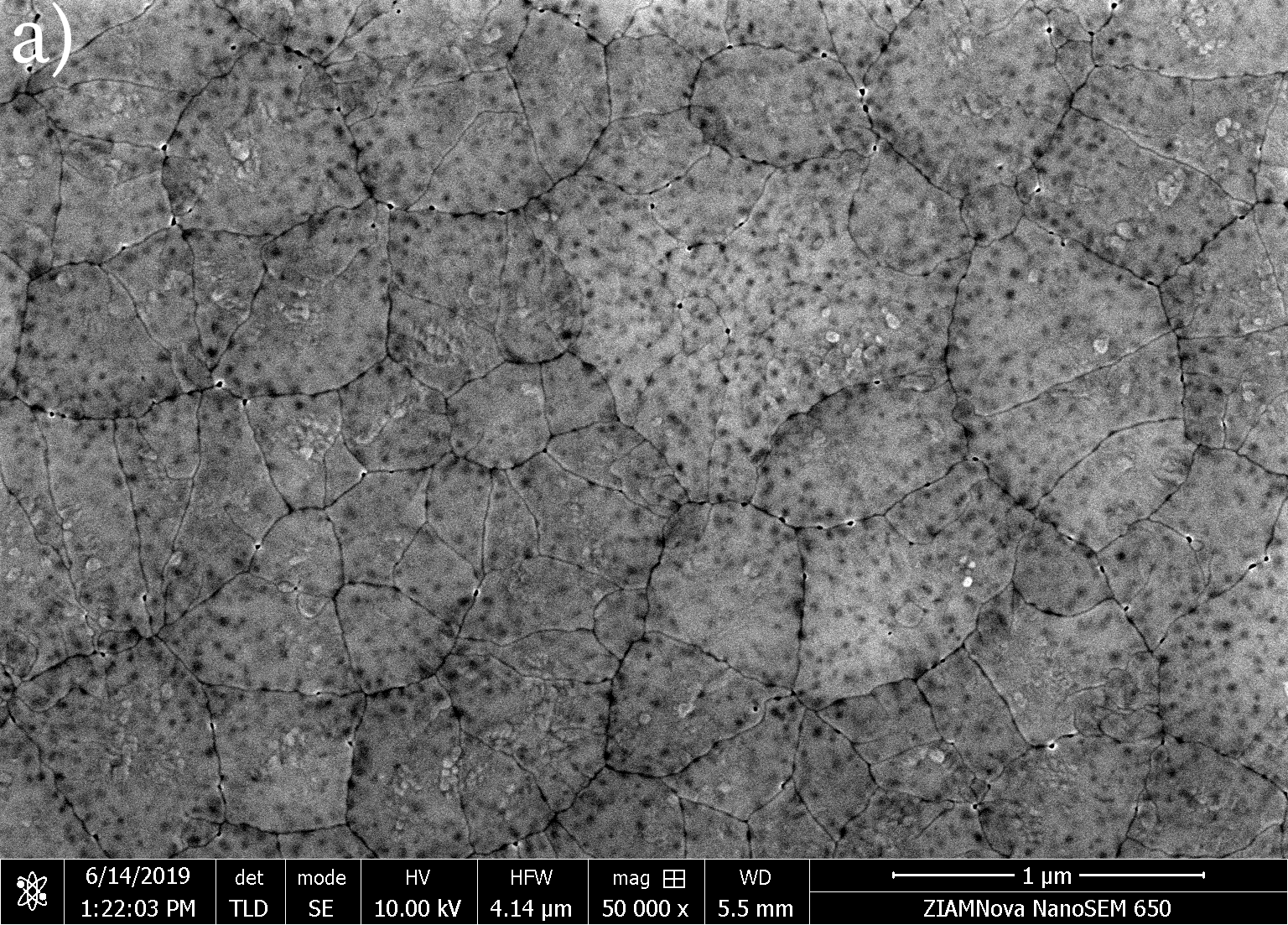}
	\refstepcounter{subfigure}
	\label{fig:sem1}
\end{subfigure}
~
\begin{subfigure}[h]{\columnwidth}
\centering
	\includegraphics[width=0.7\columnwidth]{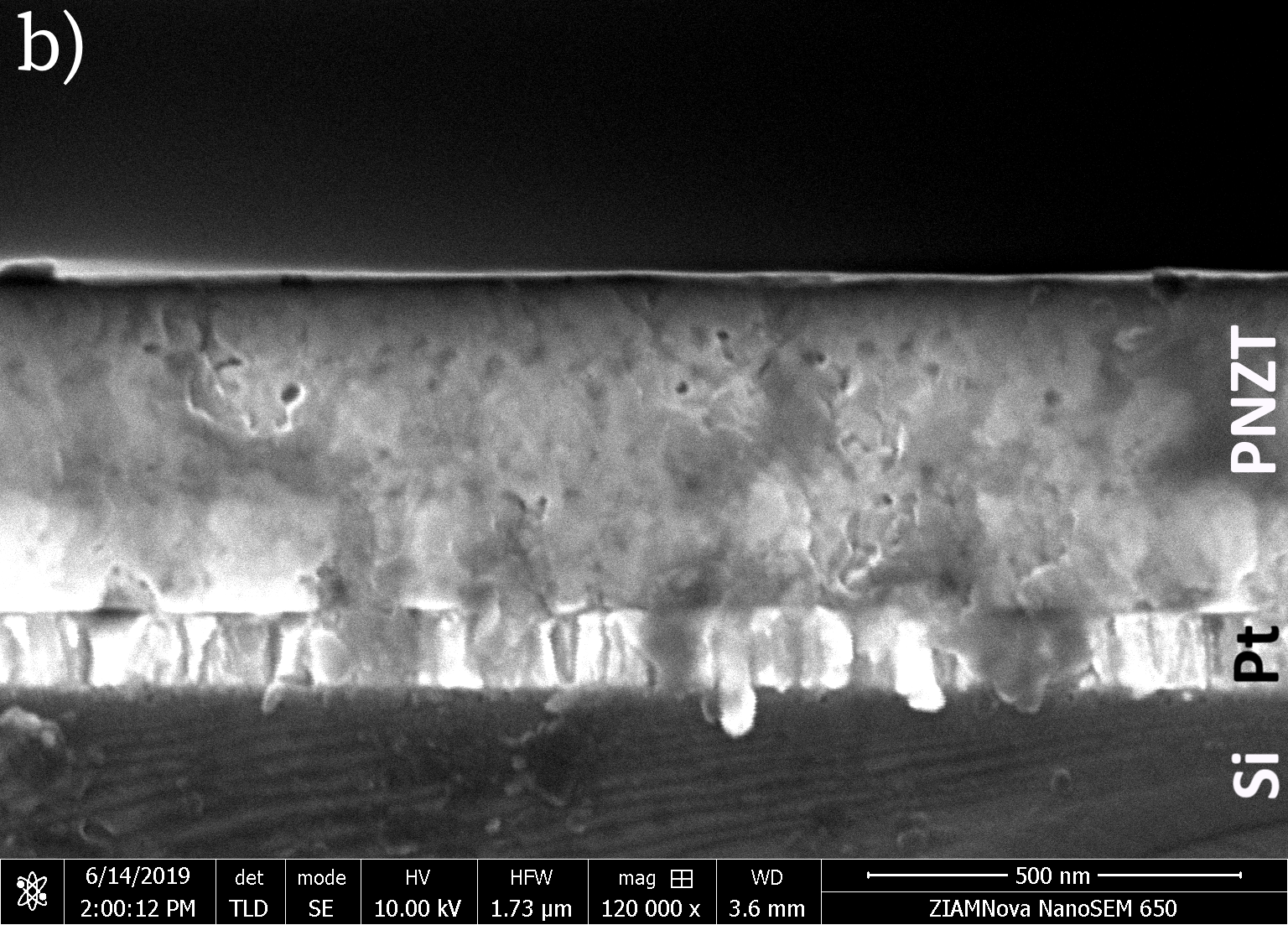}
	\refstepcounter{subfigure}
	\label{fig:sem2}
\end{subfigure}
\caption{(a) Plan-view and (b) cross-section SEM images of a nine-layer PNZT thin film stack with a PbO overcoat. The total thickness of the film is 440 nm.}
\label{fig:sem}
\end{figure}

To achieve such control, an alternative method was used here. A relatively small excess of lead of 10\% was added to the sol, compensating for diffusion but not evaporation. Additionally, a layer of pure lead oxide sol was deposited before the final pyrolysis step, compensating for evaporation from the film surface (see ref. \citenum{Brennecka2010}). The resulting film shows a dense structure with few voids and grains with sizes from several hundred nanometers up to 1 micrometer (figure \ref{fig:sem}). Using this procedure, no lead-deficient pyrochlore phase was found and no cracks or leakage paths are visible. These observations show that the combination of a lead excess in the sol with a lead oxide overcoat is effective at producing high quality thin films.  A columnar grain structure is commonly observed in PZT thin films derived from traditional sol-gel methods based on 2-methoxyethanol, due to bottom-up growth of the grains after heterogeneous nucleation at the film-electrode interface. Such structure is not present in these films (figure \ref{fig:sem2}), indicating more homogeneous nucleation. This may be the result of the high organic content of the as-deposited films compared to traditional sol-gel-derived films.

\begin{figure}[ht]
\centering
\includegraphics[width=0.7\columnwidth]{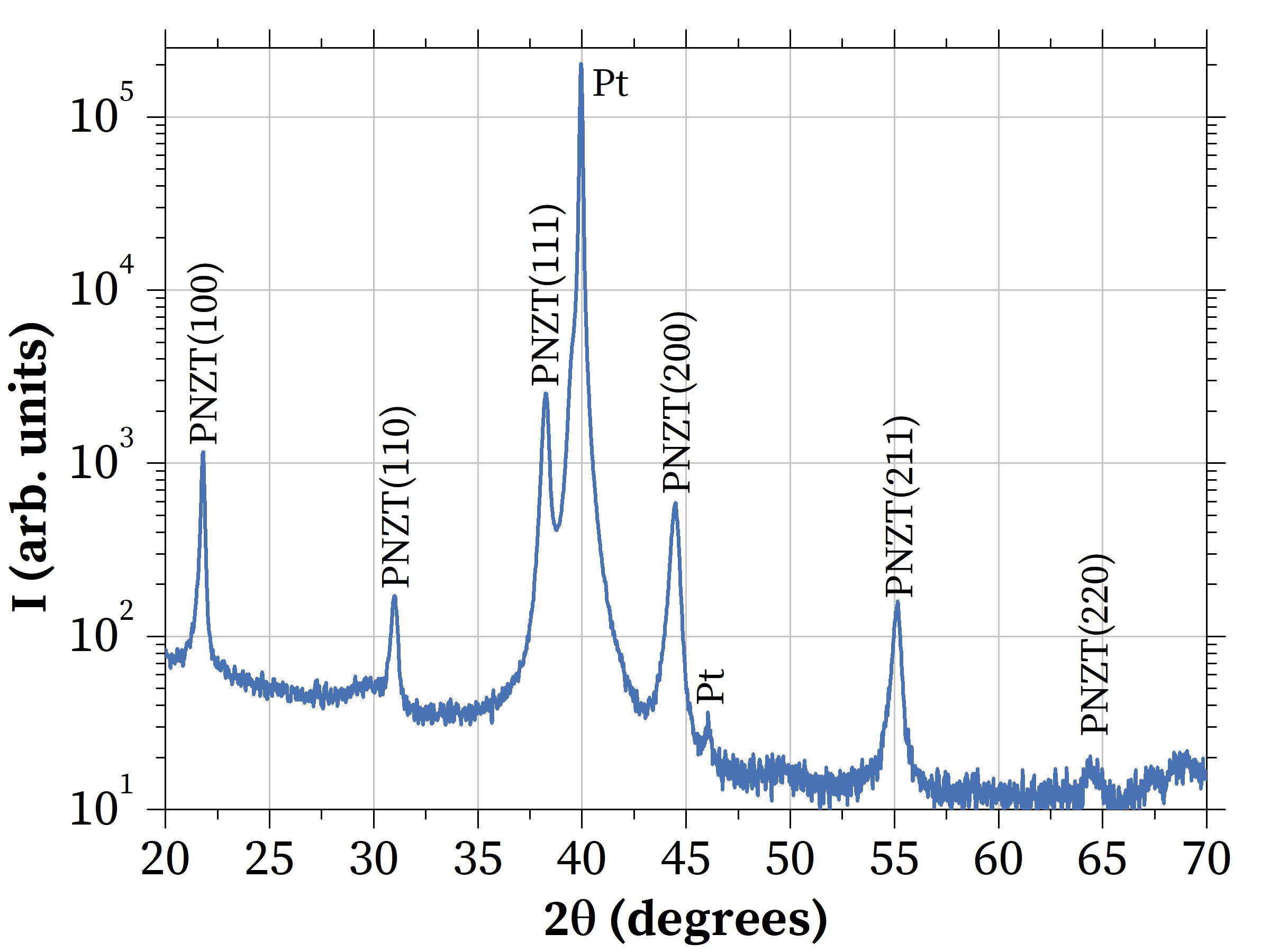}
\caption{XRD pattern of the nine-layer thin film stack.}
\label{fig:xrd1}
\end{figure}

An x-ray diffraction pattern of the same film is shown in figure \ref{fig:xrd1}. The pattern shows a pure perovskite PNZT phase with no impurity peaks, except those originating from the platinized silicon substrate. No peak splitting is observed due to the broadening of the peaks. A small Pt(200) peak is present due to the top electrode, which is not perfectly (111) oriented. The preferential orientation of the PNZT film can be quantified by normalizing the integrated peak intensities with the intensities of the x-ray diffraction patterns of a powdered sample using the following expression:\cite{Balma2014}

\begin{equation}
\label{eq:preforient}
P(h_ik_il_i) = \frac{\dfrac{I(h_ik_il_i)}{I^*(h_ik_il_i)}}{\displaystyle \sum\limits_{hkl}\dfrac{I(hkl)}{I^*(hkl)}}
\end{equation}

where $P(h_ik_il_i)$ is a texture index quantifying the preferred orientation of the sample, $I(h_ik_il_i)$ is the intensity in the thin film sample and $I^*(h_ik_il_i)$ is the intensity in the powdered sample. The values in table \ref{tab:preforient} were obtained using the data in figure \ref{fig:xrd1}.

\begin{table}[ht]
\caption{Texture index values of the thin film sample}
\label{tab:preforient}
\centering
\begin{tabular}{l l}
$<h_ik_il_i>$ & $P(h_ik_il_i)$ \\
\hline
<100> & 0.25\\
<110> & 0.0086\\
<111> & 0.59\\
<200> & 0.11\\
<211> & 0.020\\
<220> & 0.016\\
\hline
\end{tabular}
\end{table}

A <111> orientation is preferred in these films due to the <111> texture of the underlying platinum electrode, showing that at least some of the film nucleates heterogeneously at the film-electrode interface. However, it is evident that some of the film nucleates homogeneously, resulting in a decreased <111> texture of the film. This is in agreement with the lack of columnar grains in the film.

\begin{figure}[ht]
\centering
\begin{subfigure}[h]{\columnwidth}
\centering
	\includegraphics[width=0.7\columnwidth]{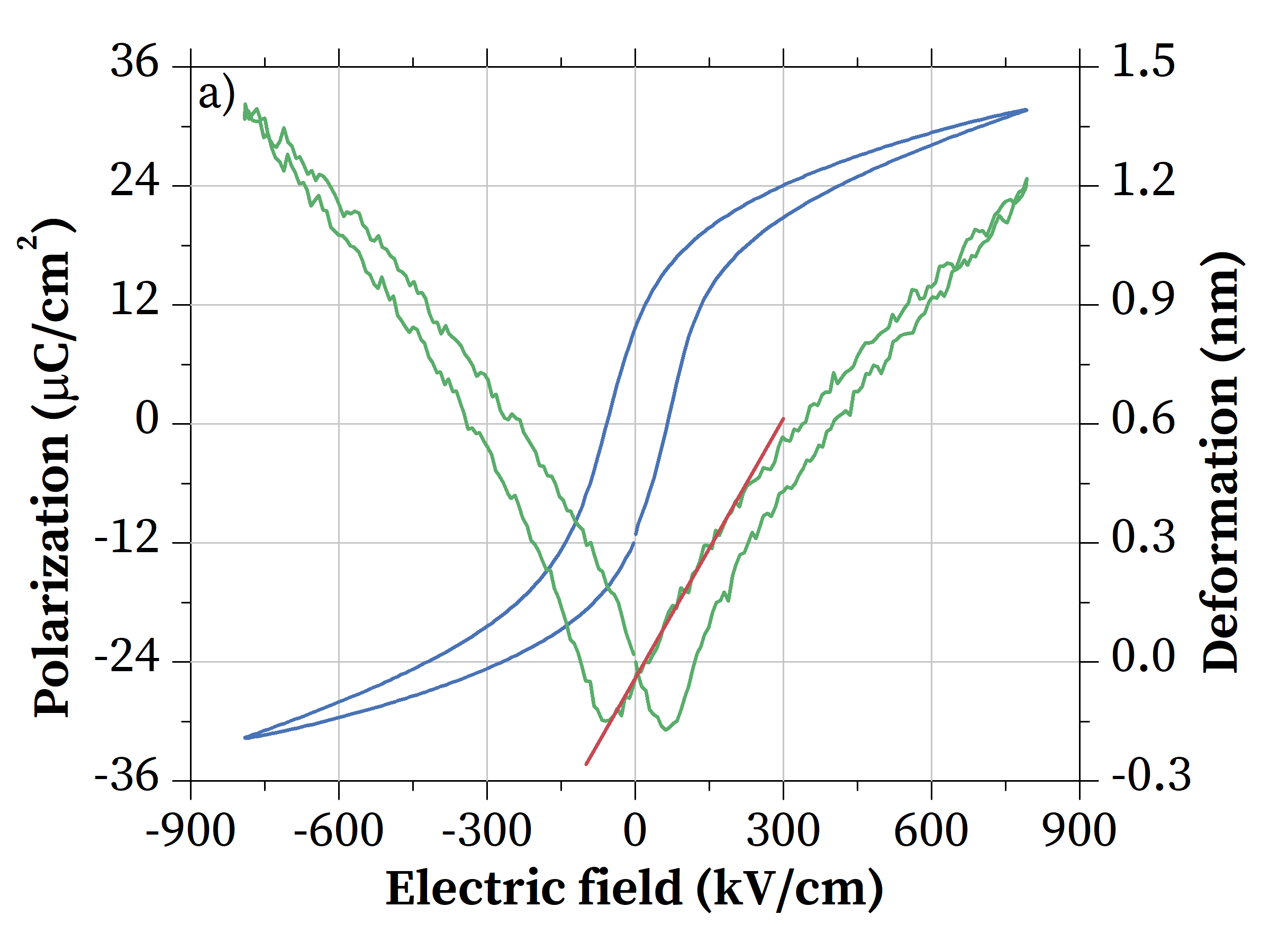}
	\refstepcounter{subfigure}
	\label{fig:piezofilmfinal}
\end{subfigure}
~
\begin{subfigure}[h]{\columnwidth}
\centering
	\includegraphics[width=0.7\columnwidth]{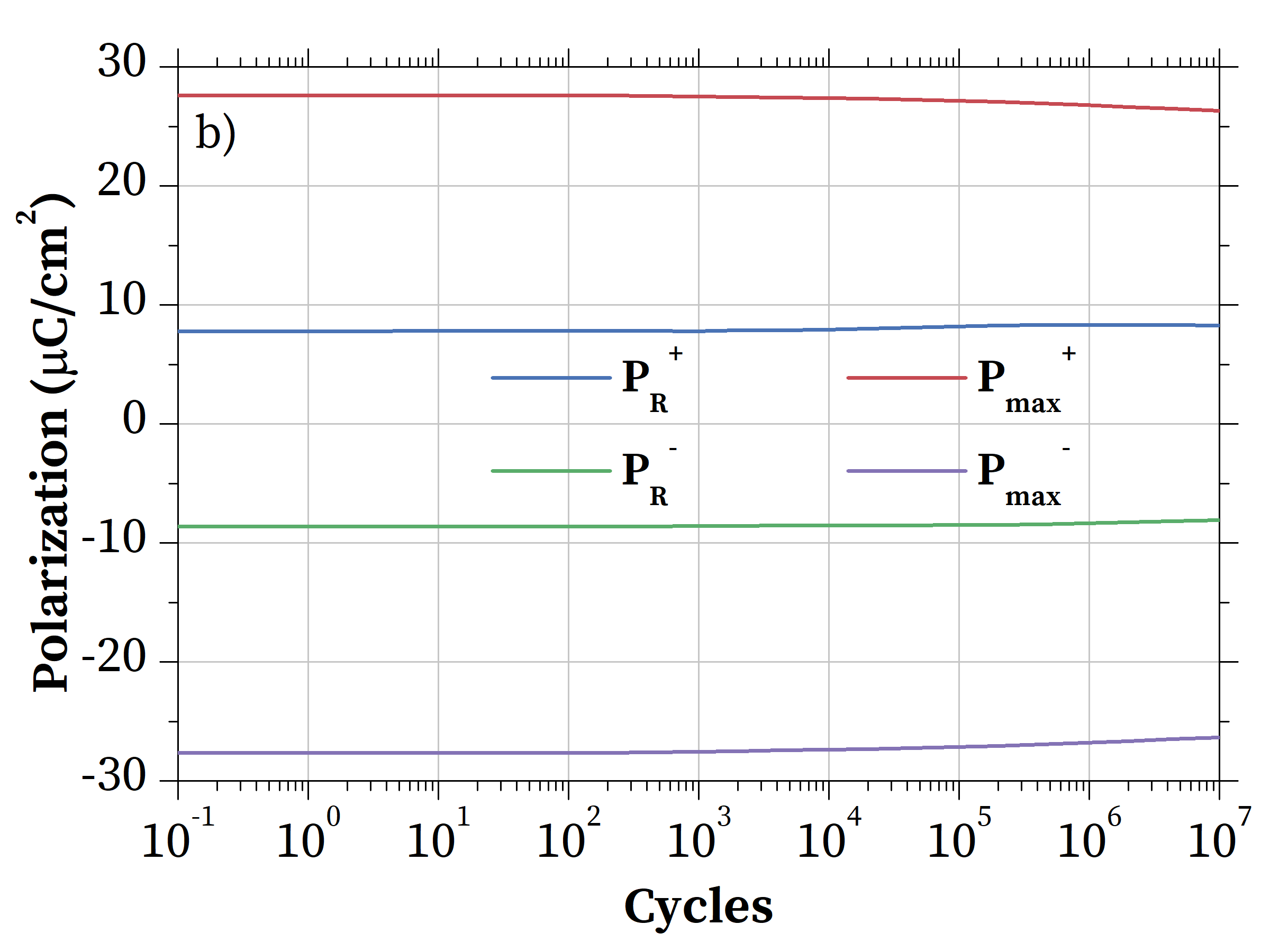}
	\refstepcounter{subfigure}
	\label{fig:fatiguefilmfinal}
\end{subfigure}
\caption{(a) Ferroelectric and strain loop of the nine-layer PNZT thin film stack. The longitudinal piezoelectric coefficient is obtained from the slope of the strain loop at zero electric field, as indicated by the red tangent line. (b) Fatigue response of the film up to $\mathrm{10^7}$ cycles.}
\label{fig:piezofilm}
\end{figure}

Figure \ref{fig:piezofilmfinal} shows the ferroelectric hysteresis loop and strain loop of a nine-layer PNZT thin film stack.
A double-beam laser interferometer, which corrects for substrate bending to extract the true deformation of the film, was used to collect these loops. The loops were collected by sweeping the potential applied to the top electrode between +800 kV/cm and --800 kV/cm using a triangular waveform at a frequency of 100 Hz. The longitudinal piezoelectric coefficient is extracted from the strain loop by determining its slope at zero applied field. The film showed a remnant polarization of 10.5 $\mathrm{\mu C/cm^2}$, an average coercive field of 61.3 kV/cm, a longitudinal piezoelectric coefficient of 50 pm/V and a maximum deformation of 1.41 nm, that is 0.3\% of the thickness of the film. These values are again compared to those found in the literature, see table \ref{tab:piezocompfilm}, which displays the wide range of piezoelectric parameter values reported depending on the synthesis technique. Our piezoelectric coefficient is on the low end of this range, but improvements can likely be made. For example, fabrication of thicker films will improve piezoelectric behavior due to reduced clamping from the substrate.  

\begin{table}[ht]
    \centering
    \begin{tabular}{l l l}
        $d_{33}$ \textit{(pm/V)} & \textit{Production method} & \textit{Reference} \\
        \hline
        50 & CSD & This work\\
        \hline
        50 & Sol-gel & \citet{Balma2014}\\
        77 & Sol-gel & \citet{Taylor2000}\\
        200 & OMCVD & \citet{Lefki1994}\\
        400 & Sol-gel & \citet{Lefki1994}\\
        85 & Sol-gel & \citet{Ledermann2003}\\
        70-80 & Sol-gel & \citet{Zavala1997}\\
        200 & Sol-gel & \citet{Chen1996}\\
        25 & Sol-gel & \citet{Wang2002}\\
        57.6 & Sol-gel & \citet{Lian2000}\\
        164* & PLD & \citet{Nguyen2014}\\
        106 & PLD & \citet{Goh2005}\\
        \hline
    \end{tabular}
    \caption{Comparison of the longitudinal piezoelectric coefficient of the PNZT thin film fabricated using the ethylene glycol CSD method with literature values. These results are for undoped PZT, unless otherwise noted. *: 1 \% Nb doping, OMCVD = organometalic chemical vapor deposition, PLD = pulsed laser deposition.}
    \label{tab:piezocompfilm}
\end{table}

Figure \ref{fig:fatiguefilmfinal} shows the fatigue response of the film. The film was switched at a frequency of 200 Hz with an electric field amplitude of 114 kV/cm, that is above the coercive field. Ferroelectric hysteresis loops were collected at 3 points/decade with a field amplitude of 800 kV/cm and a frequency of 100 Hz. The film is stable to fatigue for at least $\mathrm{10^7}$ cycles. These results are similar to those reported in the literature (see, e.g., refs. \citenum{Balma2014, Klissurska1997}).

To summarize, a sol-gel method was developed based on ethylene glycol as a solvent and bridging ligand. This sol was used for the production of pellets and thin films of ferroelectric niobium-doped lead zirconate titanate. This sol offers the advantages of a lower toxicity solvent, improved stability during storage, decreased sensitivity to atmospheric moisture and the applicability to the synthesis of both bulk and thin film products. Furthermore, the synthesis of the sol is less complex than that of traditional, 2-methoxyethanol-based sols. DTA of the sol shows that decomposition of the gel is finished at 400\textcelsius, with crystallization of the desired PNZT phase occurring at higher temperatures.  
Pellets of bulk PNZT were produced, having a density of 86.9\% of the theoretical density and a small lead oxide impurity. These show good properties with a coercive field of 68 kV/cm, a remnant polarization of 9.5 $\mathrm{\mu C/cm^2}$ and a piezoelectric coefficient of 441 pm/V, in line with literature values for similar PZT compositions.\cite{Damjanovic1999} 
In addition, a nine-layer stack of PNZT thin films was fabricated from the sol by spin-coating with a thickness of 440 nm. An excess of lead was supplied to the thin films to compensate for evaporation and diffusion by combining the addition of an excess of lead precursor to the sol and the application of an overcoat of pure lead oxide. This method proved effective at suppressing the appearance of lead-deficient phases or voids in the stack. The final stack shows a dense perovskite grain structure with a weak (111) out-of-plane texture. Ferroelectric and piezoelectric characterization of the film shows ferroelectric coefficients close to literature values for thin films, with a remnant polarization of 10.5 $\mathrm{\mu C/cm^2}$, a coercive field of 61.3 kV/cm, a piezoelectric coefficient of 50 pm/V and a maximum deformation of 0.3\% of the thickness of the film. Furthermore, the film shows good stability to fatigue up to $\mathrm{10^7}$ cycles. This sol-gel method provides a safer, more water-stable alternative to traditional sol-gel methods based on 2-methoxyethanol for the fabrication of bulk and thin film products.

\section{Materials and Methods}
\subsection{Sol synthesis}
7.5 g of freeze dried lead acetate (\ch{Pb(CH_3COO)_2}, \ch{PbAc_2}, 23 mmol, 10 mol\% excess, $\mathrm\geq$ 99\%, Sigma Aldrich) and 9.4 mL ethylene glycol (\ch{(CH_2OH)_2}, EG) were added to a three-necked flask under a 0.5 lpm argon flow. An excess of lead acetate was used to compensate for losses due to evaporation and diffusion during the heat treatment steps. The suspension was heated to 90\textcelsius\ while stirring to dissolve the solids, then to 110\textcelsius\ to expel any remaining water from the solution. The sol was subsequently cooled to 90\textcelsius. 2.857 mL of titanium isopropoxide (\ch{Ti(OCH(CH_3)_2)_4}, 2.743 g, 9.65 mmol, 97\%, Sigma Aldrich), 0.210 mL niobium ethoxide (\ch{Nb(OCH_2CH_3)_5}, 0.266 g, 0.836 mmol, 99.95\%, Sigma Aldrich) and 4.686 mL of a 70 wt.\% solution of ziconium n-propoxide in 1-propanol (\ch{Zr(OCH_2CH_2CH_3)_4}, 3.425 g \ch{Zr(OCH_2CH_2CH_3)_4}, 10.5 mmol, Sigma Aldrich) were dissolved in 6.1 mL 1-propanol under inert atmosphere. The Ti/Nb/Zr solution was added to the lead sol slowly limiting exposure to air. Some precipitate formed upon addition. A further 15.7 mL of EG was added, yeilding 30 mL of solution at a concentration of metal ions of 1.5 M with a nominal composition of \ch{Pb_{1.1}Nb_{0.04}(Zr_{0.52}Ti_{0.48})_{0.96}O_3}. The suspension was stirred at 90\textcelsius\ until all precipitate had redissolved. The sol was cooled to room temperature and 4 vol.\% formamide (\ch{HCONH_2}, $\mathrm\geq$ 99\%, Sigma Aldrich) was added as a drying control chemical additive to limit the formation of cracks in the films.\cite{Hench1990} The sol was stored under inert atmosphere, where it is stable for at least 3 months. In air, the lifetime of the sol is shorter, but it is still stable for 1-2 weeks. A second sol was made in the same way, using a 20\% excess of lead precursor. This sol was used for the preparation of bulk PNZT (see below).

A separate lead oxide (PbO) sol was fabricated by dissolving 9.76 g of freeze dried lead acetate (\ch{Pb(CH_3COO)_2}, \ch{PbAc_2}, 23 mmol, 10 mol\% excess, $\mathrm\geq$ 99\%, Sigma Aldrich) in 30 mL of ethylene glycol (\ch{(CH_2OH)_2}, EG) while stirring to a final concentration of 1 M.

\subsection{Substrate preparation}
The substrates used for deposition of the PNZT sol were prepared from a (001) oriented silicon wafer without thermal oxide (Ted Pella) diced in 1x1 cm squares. The silicon substrates were cleaned ultrasonically in acetone, demineralized water and ethanol for ten minutes each. The substrates were subsequently blow dried using compressed air and loaded into a Kurt J. Lesker sputtering system. The substrated were \ch{O_2} plasma cleaned (0.15 mbar, 200 W, 5 min.) after which a titanium adhesion layer of 5/10 nm was DC sputtered (200 W, 0.2 nm/s) without breaking the vacuum. Subsequently, a 100 nm thick electrode of platinum was DC sputtered (200 W, 1.61 nm/s) onto the adhesion layer. The full electrode stack was annealed in a box furnace in air (450\textcelsius, 90 min., ramp rate 14.2 \textcelsius/s).

\subsection{Deposition procedure and heat treatment}
The platinized silicon substrates prepared as described above were again cleaned ultrasonically in acetone, demineralized water and ethanol for 10 minutes each. The substrates were blow dried using compressed air and UV/\ch{O_3} treated in an Ossila UV ozone cleaner to remove any residual organic contamination from the surface. The substrates were immediately placed in the center of the vacuum chuck of a spin coater. 75 $\mathrm{\mu L}$ of the 1.5 M sol was deposited onto the substrates. The spin coater was subsequently ramped up to the desired speed at 1000 rpm/s. It was held at this speed for 30 s, then slowed to a stop at 1000 rpm/s. The film was then placed on a hotplate at 230\textcelsius\ for drying. Additional layers were deposited after drying for the production of multilayer films. After the deposition of up to three layers, the films were pyrolyzed at 380\textcelsius\ on the hotplate and annealed by placing them in a preheated box furnace at 650\textcelsius\ for 10 minutes. Multilayer stacks of up to nine single deposited layers were produced, for which pyrolysis and annealing steps were performed every third layer. Multiple annealed layers are required to prevent the formation of leakage paths through the film. A 4x4 grid of circular top electrodes of 100 nm of platinum was sputter deposited onto the films using a hard mask.

\subsection{Pellet preparation}
For the preparation of bulk pellets of PNZT, 10 mL of the sol without an excess of lead was stirred and heated at 230\textcelsius\ on a hotplate. Some of the resulting gel was used for thermogravimetric analysis (TGA) and differential thermal analysis (DTA). The remaining gel was heated in a box furnace to 420\textcelsius\ to remove the organic groups. The resulting powder was ground using a pestle and mortar and heated again at 450\textcelsius\ for 30 minutes. The amorphous PNZT powder was pressed into 10 mm pellets under a load 6.4 ton/cm$\mathrm{^2}$. These pellets were sintered in a box furnace at 800\textcelsius\ or 1200\textcelsius\ for two hours. 

\subsection{Characterization} 
Grain structures of the films and pellets and film thicknesses were studied using an FEI Nova Nano\-SEM 650 scanning electron microscope.  X-ray diffraction data was collected using a PanAnalytical X'Pert Pro MRD or a Bruker D8 Advance diffractometer (both in Bragg-Brentano geometry) for the films and pellets respectively. DTA-TGA data was collected in argon from 200\textcelsius\ to 1200\textcelsius\ at a heating rate of 10\textcelsius/minute using a TA instruments SDT 2960 differential scanning calorimeter. Finally, ferroelectric and piezoelectric properties of the films and pellets were measured using a state-of-the-art AixACCT TF analyzer 2000 ferroelectric-piezoelectric characterization system with an AixACCT double beam (films) or a Sios single beam (pellets) interferometer. The use of a double beam interferometer eliminates the contribution of the bending of the substrate to the measured deformation of the film.

\begin{acknowledgement}

We gratefully acknowledge the invaluable help of Jacob Baas and Henk Bonder in the lab. M.A. acknowledges financial support of a FOM-f Fellowship of the Dutch Research Council (NWO). 

\end{acknowledgement}

\begin{suppinfo}

A listing of the contents of each file supplied as Supporting Information
should be included. For instructions on what should be included in the
Supporting Information as well as how to prepare this material for
publications, refer to the journal's Instructions for Authors.

The following files are available free of charge.
\begin{itemize}
  \item Filename: brief description
  \item Filename: brief description
\end{itemize}

\end{suppinfo}


\bibliography{references}

\end{document}